# CMOS based image cytometry for detection of phytoplankton in ballast water

J. M. Pérez[a], M. Jofre[a], P. Martínez[a], M. A. Yáñez[b], V. Catalan[b], A. Parker[c], M. Veldhuis[d], V. Pruneri[a,e*]

[a]*ICFO – Institut de Ciencies Fotoniques, The Barcelona Institute of Science and Technology, 08860 Castelldefels (Barcelona), Spain*
[b]*LABAQUA, S.A., c) Del Dracma, 16 18, 0314, Alicante, Spain*
[c]*MemTeq Ventures Ltd. The Auction Houses, Stokesley North Yorkshire, TS9 7AB England, UK*
[d]*Marine Eco Analytics, Haventerrein 1-A, 1779 GS Den Oever, The Netherlands*
[e]*ICREA – Institució Catalana de Recerca I Estudis Avançats, Barcelona, Spain.*

ABSTRACT

We introduce an image cytometer (I-CYT) for the analysis of phytoplankton in fresh and marine water environments. A linear quantification of cell numbers was observed covering several orders of magnitude using cultures of *Tetraselmis* and *Nannochloropsis* measured by autofluorescence in a laboratory environment. We assessed the functionality of the system outside the laboratory by phytoplankton quantification of samples taken from a marine water environment (Dutch Wadden Sea, The Netherlands) and a fresh water environment (Lake Ijssel, The Netherlands). The I-CYT was also employed to study the effects of two ballast water treatment systems (BWTS), based on chlorine electrolysis and UV sterilization, with the analysis including the vitality or the phytoplankton. For comparative study and benchmarking of the I-CYT, a standard flow cytometer was used. Our results prove a limit of detection (LOD) of 10 cells/ml with an accuracy between 0.7 and 0.5 log, and a correlation of 88.29% in quantification and 96.21% in vitality, with respect to the flow cytometry results.

## 1. Introduction

Globalization has become a primary driver of one of the most prevalent forms of environmental degradation: marine invasive species; as trade continues to flourish, bio-invasion is becoming more difficult to handle (Kannan, 2015). Among the marine invasive species, microorganisms carried in ballast water (BW) can easily spread into a new habitat. This can generate a potentially devastating impact threatening the ecosystem and human activities (Moreno-Andrés et al., 2016). Of the microorganisms some species of phytoplankton can cause illness to mammals, fish, corals and other marine organisms.

BW refers to the water in the tanks of ships used to increase their stability, which is discharged into the ocean after long journeys, but it is also introducing numerous non-indigenous organisms to new ecosystems (Bax et al., 2003). BW on ships is considered as the most important vector in dispersing invasive species throughout the world (Seebens et al., 2013) as more than 150.000 metric tons of fresh/marine water can be pumped in or out in only one ballast / de-ballast operation (Dunstan and Bax, 2008). In response to the threats from continued introductions of aquatic invasive species, the United Nations - International Marine Organization (IMO) adopted the International Convention for the Control and Management of Ships' Ballast Water and Sediments (IMO, 2004). The IMO regulation sets procedures to discharge BW in ports, to minimize the spread of invasive and pathogenic organisms. Its compliance requires testing for phytoplankton, zooplankton, toxicogenic *Vibrio cholera*, *Escherichia coli* and intestinal *Enterococci* upon discharge of the ballast water in the harbor.

BW treatment systems (BWTS) represent a way of disinfecting ballast water in order to reduce the number of organisms to low risk levels for the ecosystem and human health; BWTS are either on board or port-based systems which are able to clean all BW before it is released into the harbor (Rivas-Hermann et al., 2015). The main on board and port-based treatment technologies used today are ultraviolet (UV) sterilization (Stehouwer et al., 2015) and chlorine electrolysis (Maranda et al., 2013).

Collaborative efforts between biologists, physicists, engineers, chemists and material researchers have yielded novel strategies in understanding complex marine ecosystems. A variety of analytical methods have been used to identify changes in populations of marine organisms ranging from large to small scale like: remote spectrometry from satellites and airplanes, in situ spectrometry, (laser)-induced fluorescence, microscopy and flow cytometry (Golden et al., 2012a).

Other biosensing systems for on-board analysis of BW and quantification of the living organisms have been developed and reported. For example, sensors based on molecular and genetic engineering methods (Sanchez-Ferandin et al., 2013; Wollschläger et al., 2014); others relying on the photosynthetic properties, universally present in phytoplankton (Golden et al., 2012b; Meneely et al., 2013). In a previous paper, we introduced an optical reader based on angular spatial frequency processing and incorporating consumer electronics complementary-metal-oxide semiconductor (CMOS) image sensor array for the detection of waterborne microorganisms (Perez et al., 2015).

By leveraging such optical reader, we present in this paper a field-portable image cytometry system (I-CYT) for the rapid detection and quantification of phytoplankton. The I-CYT was applied to analyze the effects of UV and chlorine BWTS.

## 2. Materials and methods

### 2.1 Image cytometer

The I-CYT is an opto-mechanical reader comprised of a CMOS image sensor array as a detector and a collimated bandlimited light emitting diode (LED) source centered at an excitation wavelength of 466nm. The collimated beam illuminates the sample volume which is contained in a disposable Poly (methyl methacrylate) (PMMA) cuvette with a capacity of up to 3ml. An interference optical filter allows the simultaneous detection of two fluorescent channels centered at 512nm and 630nm, respectively. Phytoplankton species exhibit auto-fluorescence in wavelengths above 610nm (red fluorescence), because of their chlorophyll. This fluorescence can be used for quantification but also measurements of the vitality.

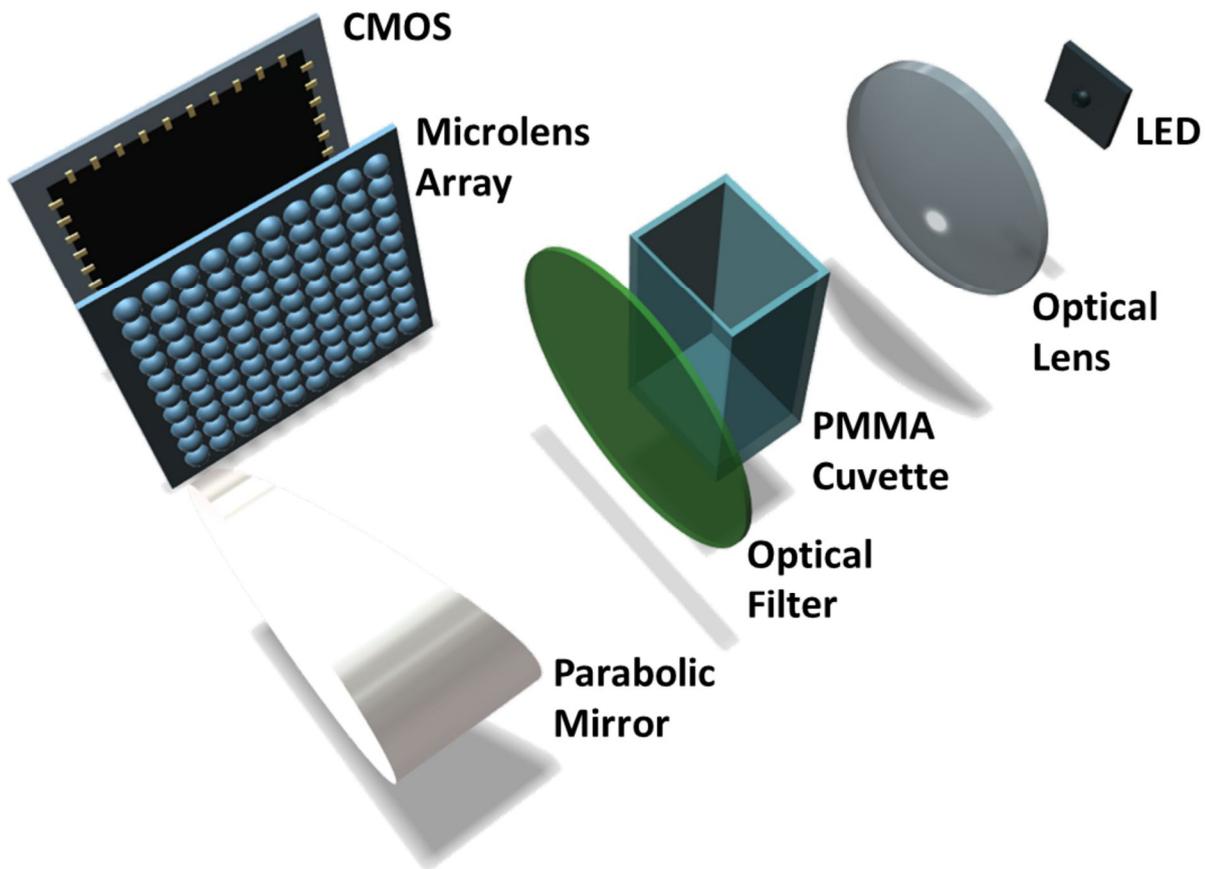

Fig. 1: Schematic of the opto-mechanics. The system is composed of a light emitting diode (LED) light source, an optical lens for collimation, an optical filter to absorb the excitation pump and select two fluorescent emission channels centred at 512nm and 630nm, respectively, a parabolic mirror which acts as an optical transforming element to propagate the optical Fourier transform of the sample towards a microlens array, which spatially samples the incoming beam, and a CMOS image sensor array that captures the light sample to process.

A parabolic mirror, placed after the filter, acts as an optical transforming element, collecting the light field from the sample and projecting its optical signal after Fourier transformation onto an array of microlenses. The microlenses physically sample the incoming beam and focus it onto different areas on the CMOS image sensor array. The combination of the optical transforming element and the microlenses array allows for the detection of the organisms in the sample in such way that they can be counted and discriminated by size. The complete I-CYT platform combines fluorescence detection with Fourier optics for a complete analysis of the sample in terms of concentration, vitality and size. Figure 1 shows a schematic of the opto-mechanics and the detected image for a reference (blank or buffer solution) and a particulate sample.

The CMOS image sensor array has a sigmoidal response to the fluorescence intensity; i.e. it has a sub-saturation region for the concentrations near the LOD, a saturation level at high concentrations and a linear region in between. The I-CYT system combines the CMOS response with a high dynamic range (HDR) capturing algorithm, thus allowing concentration estimates differing as much as 50dB.

### 2.3 I-CYT testing procedure

The dynamic range of the I-CYT was determined by means of serial dilutions of two phytoplankton species: *Tetraselmis* (14μm in cell diameter) and *Nannochloropsis* (5μm in cell diameter).

The next step was testing the functionality of the system by measuring and quantifying phytoplankton species in samples from fresh and marine waters. Furthermore, the ability of I-CYT was examined using water samples collected from full scale BWTS by passing these water samples subjected to UV sterilization or chlorine electrolysis. Phytoplankton numbers were quantified and the vitality of the cells was measured before and after the treatment.

For comparison, field samples were analyzed also with a standard flow cytometer (Beckman Coulter EPIC-XL-MCL) (Veldhuis and Kraay, 2000). The vitality of the phytoplankton was measured as the efficiency of the photosynthetic system of the phytoplankton (Schreiber, 1998). For this analysis the WALZ-Water-PAM was used measuring bulk fluorescence properties of the phytoplankton (Veldhuis et al., 2006).

The semi-quantification of organisms in cells per ml is achieved by transforming the fluorescence intensity with a 4 parameter logistic (4PL) regression (O'Connell et al., 1993), see equation 1.

$$Concentration\left[\frac{cells}{ml}\right] = C \cdot \left(\frac{A-D}{I_{CYT}-D}\right)^{1/B} \quad (1)$$

Where, (A, B, C, D) refer to the four parameters of the regression, with values ($2.17 \times 10^{-8}$, $1.31 \times 10^3$, $3.33 \times 10^7$, $4 \times 10^{-2}$) respectively, and $I_{CYT}$ refers to the fluorescence intensity. The use of a 4PL regression describes our biosensing system more suitably than a linear regression. The model has a maximum (D) and a minimum (A) built into the equation, which better describes biological systems. Parameters (C) and (B) act as offset and slope values respectively.

For the vitality index measure, the light source in the system was controlled by a pulse width modulation (PWM) signal that allowed for the detection of the minimum and maximum fluorescence of the organisms. For the maximum fluorescence ($F_m$) the sample was excited by a PWM signal with 50% duty cycle and 10μs period; for the minimum fluorescence ($F_0$), the sample was excited by a PWM signal with 50% duty cycle and 50ns period. The sample was captured under both conditions and the values of $F_m$ and $F_0$. The vitality index was calculated as follows:

$$Vitality = 1 - \frac{F_0}{F_m} \quad (2)$$

## 2.2 Sample collection and preparation

Controlled cultures of *Tetraselmis* and *Nannochloropsis* were used for laboratory test and system calibration. Both *Tetraselmis* and *Nannochlorpsis* were taken from concentrated stocks of green microalgae, the former of the *Tetraselmis chuii* species and the latter of the *Nannochloropsis oculata* species; both purchased from Acuinuga (A Coruña, ES). For the system validation in a BW environment, samples of both marine and fresh water were measured; the marine and fresh water samples were subjected to BWTS chlorine electrolysis; the fresh water samples were also exposed to BTWS UV sterilization.

Dilutions of the samples (1/10 v/v) were made in a marine water medium filtered under 0.2μm with a hollow membrane filter CellTrap™. The *Tetraselmis* culture media contained high levels of debris, especially at the highest concentration measured (1/10²). To avoid the absorption of excitation light from said debris material, a cell extraction protocol was performed to separate the *Tetraselmis* cells at this concentration. The protocol consisted on the centrifugation of the sample at 1500 revolutions per minute (rpm) for 10 minutes at 4ºC; which would effectively separate the larger debris from the *Tetraselmis* cells. After centrifugation the supernatant was recovered and measured with the I-CYT.

Fresh water samples were collected from Lake Ijssel (The Netherlands) and marine water samples from brackish water off the coast of Den Oever (Dutch Wadden Sea, The Netherlands). One of the fresh water samples was concentrated from its original volume of 400ml down to 3ml using CellTrap™, in this case the $I_{CYT}$ intensity recorded by the reader was corrected using the eluted volume (3ml taken (eluted) from the filter), original volume (400ml originally sampled) and recovery rate (RR). The recovery rate of the CellTrap™ (RR) is 0.98 as reported by the manufacturer.

In the case of the UV BWTS, the treatment included two different steps. The first consisted in exposing the water sample to UV light followed by a

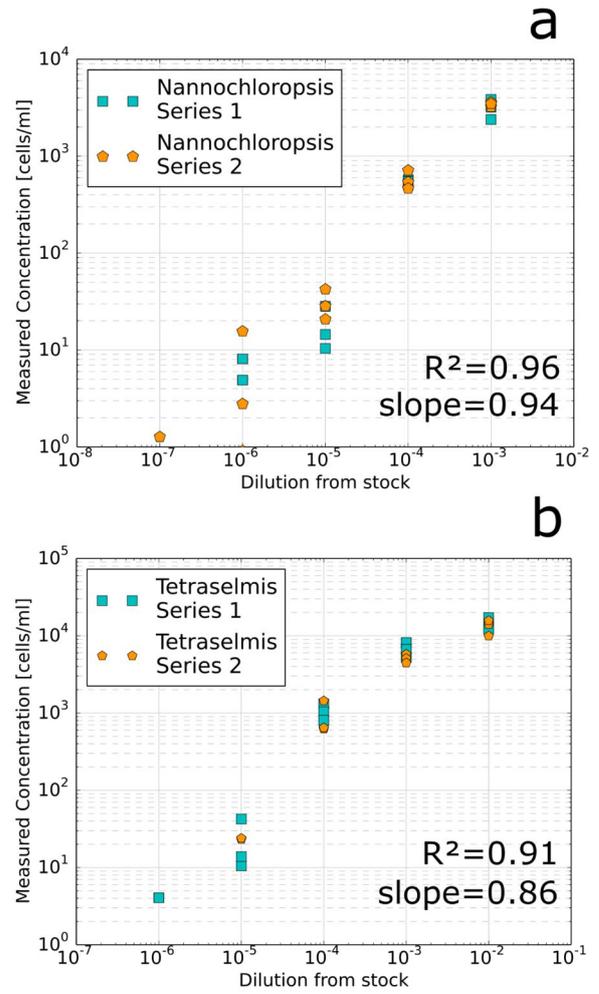

**Figure 2:** Results for the *Tetraselmis* and *Nannochloropsis* samples. (a) *Nannochloropsis* results, with concentrations measured by the I-CYT from $3 \cdot 10^3$ cells/ml at a dilution from the stock of $10^{-3}$, to <10 cells/ml at a dilution from the stock of $10^{-7}$. (b) *Tetraselmis* results, with concentrations measured by the I-CYT from $10^4$ cells/ml at a dilution from the stock of $10^{-2}$, to <10 cells/ml at a dilution from the stock of $10^{-6}$.

| | DILUTION FROM STOCK | INTRA-ASSAY DEVIATION | | INTER-ASSAY DEVIATION |
|---|---|---|---|---|
| | | Series 1 | Series 2 | |
| **NANNOCHLOROPSIS** | 10^-3 | 0.08 | 0.01 | 0.06 |
| | 10^-4 | 0.02 | 0.07 | 0.05 |
| | 10^-5 | 0.18 | 0.12 | 0.20 |
| | 10^-6 | 0.11 | 0.51 | 0.41 |
| | 10^-7 | 0 | 0.59 | 0.47 |
| **TETRASELMIS** | 10^-2 | 0.06 | 0.08 | 0.07 |
| | 10^-3 | 0.07 | 0.04 | 0.08 |
| | 10^-4 | 0.08 | 0.16 | 0.14 |
| | 10^-5 | 0.26 | 1.11 | 1.32 |
| | 10^-6 | 1.91 | 0.18 | 1.49 |

**Table 1:** Intra-assay and inter-assay deviation for both species at all concentrations measured.

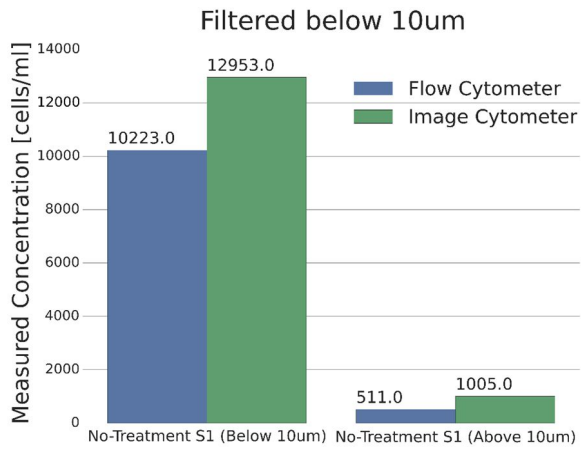

**Figure 3:** Concentration in [cells/ml] as measured by the proposed image cytometer and the reference flow cytometer. The two instruments show very similar results, with a deviation of 0.05 and 0.14 for the window below and above 10μm size, respectively.

24 hour holding; the second one included an additional UV exposure step after the 24 hour holding period.

## 3. Results and discussion

We measured unialgal cultures of *Tetraselmis* and *Nannochloropsis* to evaluate the linearity, repeatability, reproducibility and LOD of the proposed I-CYT platform. The fresh and marine water samples were measured for validation purposes in which we correlated our results with a flow cytometer standard method.

### 3.1 Detection and quantification of Tetraselmis and Nannochloropsis

The initial tests were performed in a laboratory environment, where two independent series of both *Tetraselmis* and *Nannochloropsis* were measured over five orders of magnitude in 1/10 (v/v) dilutions. Each sample was measured three times. The serial dilutions evaluated the linearity of the I-CYT and its LOD, the sets of three measurements evaluated the repeatability of the reader and the comparison between series evaluated its reproducibility.

Figure 2 compiles the results for the *Tetraselmis* and *Nannochloropsis* samples measured. Fig 2a displays the *Nannochloropsis* results, for both series and all repetitions. At the highest concentration measured ($10^{-3}$ dilution), the I-CYT reports a concentration of $3·10^3$ cells/ml; the lowest concentration detected was 6 cells/ml, at a $10^{-6}$ dilution. One more dilution was measured ($10^{-7}$), but went undetected as it was on average below the I-CYT's baseline. Fig 2b displays the results for *Tetraselmis*, of both series and all repetitions. At the highest concentration measured ($10^{-2}$ dilution), the I-CYT reports a concentration of $10^4$ cells/ml; the lowest concentration detected was 18 cells/ml, at a $10^{-5}$ dilution. One more dilution was measured ($10^{-6}$), but went undetected as it was on average below the I-CYT's baseline.

Table 1 shows the intra-assay and inter-assay logarithmic deviations for both species at all concentrations measured. On average the platform exhibits an intra-assay deviation of 0.282 and an inter-assay deviation of 0.1; this translates into an accuracy between 0.7 and 0.5 log. For comparison, when using a microscope, the *Nannochloropsis* stock showed a concentration of $7·10^6$ cells/ml, while the *Tetraselmis* stock of $10^6$ cells/ml.

To evaluate the LOD, a total of 10 independent samples per organism were measured at the lowest detected concentrations; 6 cells/ml for *Nannochloropsis* and 18 cells/ml for *Tetraselmis*, as measured with the I-CYT. The *Nannochloropsis* samples had an inter-assay deviation of 0.38; the *Tetraselmis* sampleshad an inter-assay deviation of 0.20. Deviations are taken as the standard deviation of the base 10 logarithm of the observations.

### 3.2 Marine water and fresh water analysis, and BWTS validation

Unlike cultures, samples collected from the field consists of a larger variety of phytoplankton differing in size and chlorophyll content. These samples were therefore filtered over a 10 micron net to determine the size ranges. Figure 3 shows the comparison between the concentration in [cells/ml] as measured by both platforms (image and flow cytometer) below and above the 10μm size threshold. The platforms gave very close results, with a deviation of 0.05 and 0.14 for the window below and above threshold, respectively.

Figure 4 displays the marine water samples measured before and after the disinfection step using a chlorine electrolysis BWTS. The phytoplankton population was quantified in the two windows of interest; larger (a) and smaller (b) than the 10μm threshold. The chlorine resulted in a reduction of the phytoplankton population in both regions. This can be especially noted in samples 1 and 3, where the decrease in cell numbers was one order of magnitude with respect to untreated samples. In sample 2 the BWTS has a lower impact: the reduction was only a 50% numerical reduction below the size threshold and even smaller above the 10 micron size range.

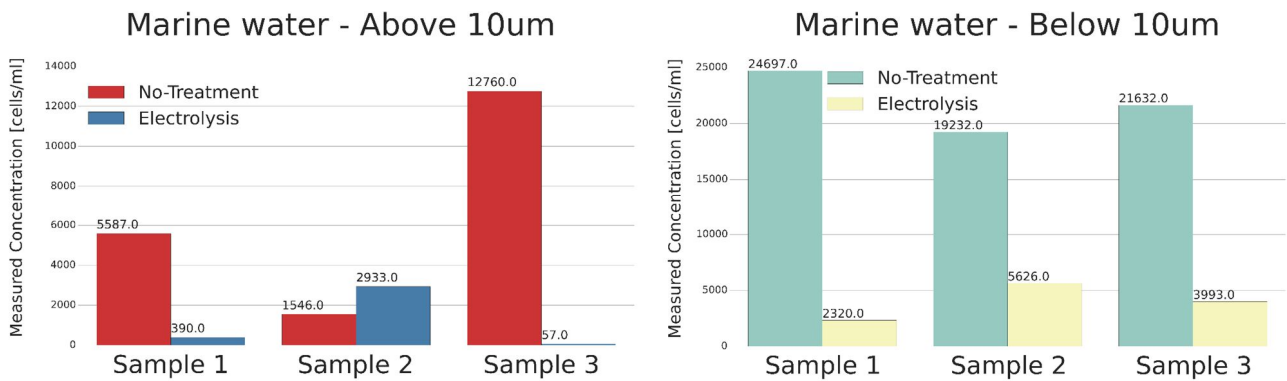

**Figure 4:** Marine water samples measured before and after a BWTS of electrolysis by chlorine. The phytoplankton population was quantified in the two windows of interest; above (a) and below (b) the 10 μm threshold. The effect of the electrolysis by chlorine, reduces the phytoplankton population in both regions. This can be specially noted in samples 1 and 3, were the decrease in concentration is of one order of magnitude. In sample 2 the system has a lower impact, were it reduced the population in half an order of magnitude below the threshold and seem to stay the same above the threshold, where it shows a slight increase within the margin of 0.5 log deviation reported in the laboratory measurements.

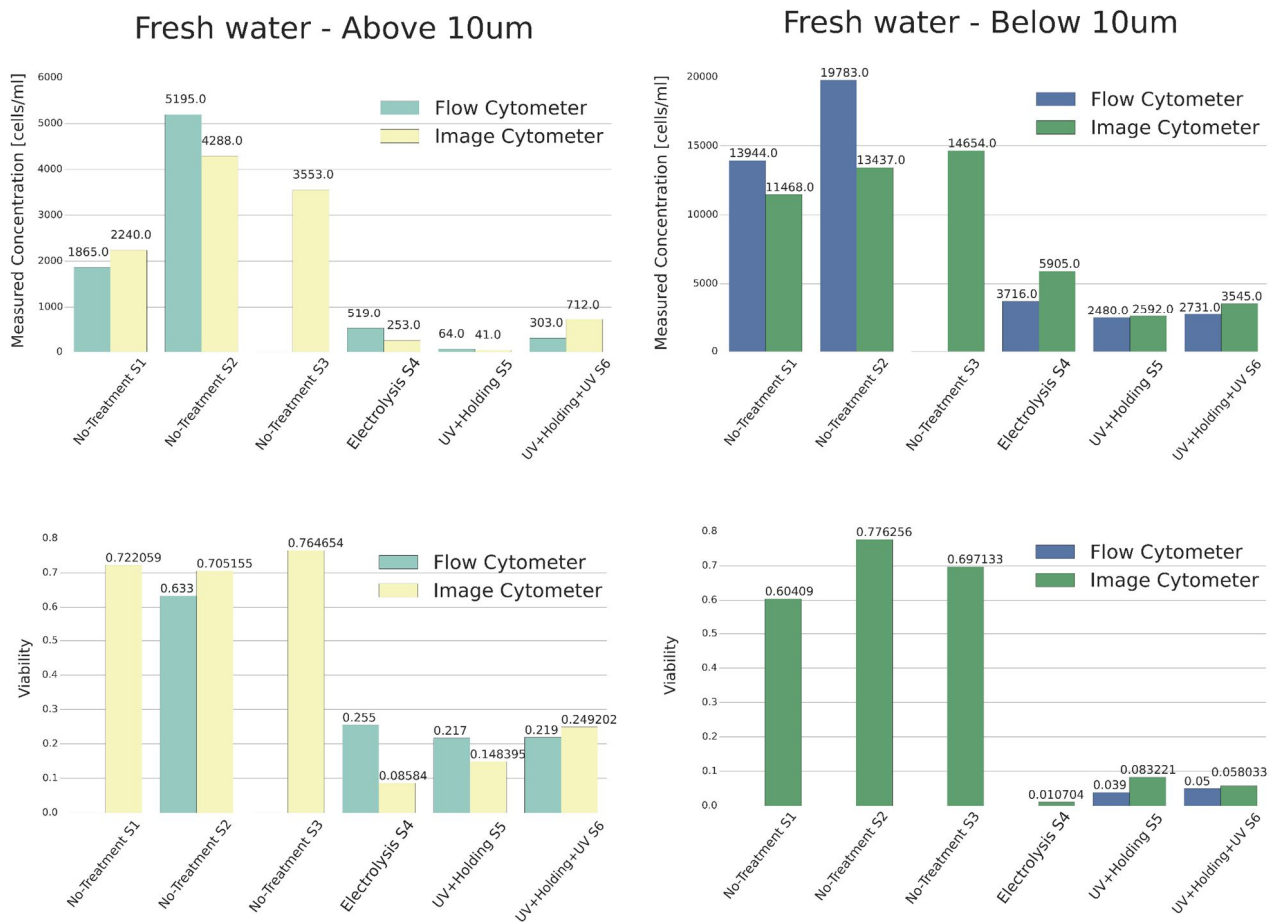

**Figure 5:** Fresh water samples were tested before and after three different protocols of BWTS (electrolysis by chlorine, UV sterilization with 1 day holding, and UV sterilization with one day holding and a second UV exposure after holding). Both phytoplankton population and viability index were measured for the two windows of interest; above (a,c), and below (b,d) 10 μm size threshold. The samples were also measured with a gold reference flow cytometer. In terms of quantification, the image cytometer has correlation factors of 94.89% and 81.70% above and below the 10 μm size threshold, respectively; in terms of viability, the correlations are 92.43% and 100% above and below size threshold, respectively. The BWTS largely affect the phytoplankton population and viability as can be seen by the results obtained with both the image and flow cytometry platforms.

Figure 5 displays the results of concentration and vitality for fresh water samples before and after three different treatment protocols of the BWTS (chlorine electrolysis, UV sterilization with 1-day holding, and UV disinfection with one day holding and a second UV exposure after holding). A total of six samples were tested; samples 1 through 3 had no treatment, and samples 4 through 6 with three different BWTS protocols. The vitality index of the non-treated samples ranged between 0.6 and 0.77, for both size ranges.

Samples 2 and 3 in Figure 5 came from the same original bulk sample, with the difference that sample 3 was concentrated from 400ml down to 3ml (CellTrap™). This shows the high performance of the combined I-CYT reader and the CellTrap™ filter in measuring concentrated samples. The deviation in concentration between samples 2 and 3 was 0.01 with the filter presenting a recovery rate of 0.96 (in line with the performance characteristics of the CellTrap™) (Figure 5 (a)), proving the efficacy of the filter and the reproducibility of the I-CYT reader.

Sample 4 in Figure 5 was treated with chlorine electrolysis, the results show a decrease in cell numbers of at least one order of magnitude (Figure 5 (a) and (b)), analog to the effects of BWTS in marine water samples (Figure 4). The vitality index of the sample after the treatment was 0.01 below the size threshold and 0.08 above it. For comparison the latter compared to 0.25 measured with PAM-fluorometry (Figure 5 (c) and (d)).

Finally, samples 5 and 6 show the results after UV sterilization and 1-day holding (sample 5), and second UV exposure after holding (sample 6). Both concentration and the vitality index were reduced to similar levels in both treatments and for all size regions, data corroborated by the flow cytometry and PAM fluorometry results. The second UV-disinfection treatment after 1 day holding time did not differ significantly from the first treatment, therefore the increased efficacy of the second treatment can be considered as minor.

## 4. Conclusions

We have proposed a new optical system, an image cytometer (I_CYT), for the analysis of phytoplankton in fresh and marine water environments. The results and their rerpoducibility demonstrated the high performance of the I-CYT for the quantification of phytoplankton, both inside and outside a laboratory environment.

In the present study the focus has also been on quantifying the efficacy of BWTS. To this end, by using pulse width modulation (PWM) of the light pump source, accurate measurements of vitality of the phytoplankton were achieved.

The experiments clearly indicate that the proposed I-CYT has comparable performance to standard flow cytometry equipment and, given its portable compact form, is a very promising solution for the analysis of BW and prevention of spreading of invasive species.

Future work will include the simultaneous use of a second fluorescence channel for the analysis and quantification of waterborne bacteria; resulting in a complete analysis of the sample according to Ballast water regulation in a single measurement.

## Acknowledgements

This work was funded by the European Union's Horizon 2020 research and innovation program under grant agreement no. 642356 (CYTOWATER project), by Fundació Privada Cellex, by the Spanish MINECO (Severo Ochoa grant SEV-2015-0522), by the European Regional Development Fund (FEDER) through grant TEC2013-46168-R, by AGAUR 2014 SGR 1623.

## References

Bax, N., Williamson, A., Aguero, M., Gonzalez, E., Geeves, W., 2003. Marine invasive alien species: a threat to global biodiversity. Mar. Policy 27, 313–323. doi:10.1016/S0308-597X(03)00041-1

Dunstan, P.K., Bax, N.J., 2008. Management of an invasive marine species: defining and testing the effectiveness of ballast-water management options using management strategy evaluation. ICES J. Mar. Sci. 65, 841–850. doi:10.1093/icesjms/fsn069

Golden, J.P., Hashemi, N., Erickson, J.S., Ligler, F.S., 2012a. A microflow cytometer for optical analysis of phytoplankton. SPIE Proc. 8212, 82120G–82120G–6. doi:10.1117/12.905679

Golden, J.P., Hashemi, N., Erickson, J.S., Ligler, F.S., 2012b. <title>A microflow cytometer for optical analysis of phytoplankton</title>. SPIE Proc. 8212, 82120G–82120G–6. doi:10.1117/12.905679

IMO, 2004. International Convention for the Control and Manangement Of Ship's Ballast Water and Sediments. IMO Doc. BWM/CONF/36 London:IMO.

Kannan, V., 2015. Globalization and government rugulations: Invasive species management in an era of interdependence. 3808 a J. Crit. Writ. 10, 8–12.

Maranda, L., Cox, A.M., Campbell, R.G., Smith, D.C., 2013. Chlorine dioxide as a treatment for ballast water to control invasive species: Shipboard testing. Mar. Pollut. Bull. 75, 76–89. doi:10.1016/j.marpolbul.2013.08.002

Meneely, J.P., Campbell, K., Greef, C., Lochhead, M.J., Elliott, C.T., 2013. Development and validation of an ultrasensitive fluorescence planar waveguide biosensor for the detection of paralytic shellfish toxins in marine algae. Biosens. Bioelectron. 41, 691–697. doi:10.1016/j.bios.2012.09.043doi:10.1016/j.pocean.2015.03.010

Moreno-Andrés, J., Romero-Martínez, L., Acevedo-Merino, A., Nebot, E., 2016. Determining disinfection efficiency on E. faecalis in saltwater by photolysis of H2O2: Implications for ballast water treatment. Chem. Eng. J. 283, 1339–1348. doi:10.1016/j.cej.2015.08.079

O'Connell, M.A., Belanger, B.A., Haaland, P.D., 1993. Calibration and assay development using the four-parameter logistic model. Chemom. Intell. Lab. Syst. 20, 97–114. doi:10.1016/0169-7439(93)80008-6

Perez, J.M., Jofre, M., Martinez, P., Yanez, M.A., Catalan, V., Pruneri, V., 2015. An image cytometer based on angular spatial frequency processing and its validation for rapid detection and quantification of waterborne microorganisms. Analyst 7734–7741. doi:10.1039/c5an01338k

Rivas-Hermann, R., Köhler, J., Scheepens, A.E., 2015. Innovation in product and services in the shipping retrofit industry: a case study of ballast water treatment systems. J. Clean. Prod. 106, 443–454. doi:10.1016/j.jclepro.2014.06.062

Sanchez-Ferandin, S., Leroy, F., Bouget, F.Y., Joux, F., 2013. A new, sensitive marine microalgal recombinant biosensor using luminescence monitoring for toxicity testing of antifouling biocides. Appl. Environ. Microbiol. 79, 631–638. doi:10.1128/AEM.02688-12

Schreiber, U., 1998. Chlorophyll fluorescence : new instruments for special applications 4253–4258.

Seebens, H., Gastner, M.T., Blasius, B., 2013. The risk of marine bioinvasion caused by global shipping. Ecol. Lett. 16, 782–90. doi:10.1111/ele.12111

Stehouwer, P.P., Buma, A., Peperzak, L., 2015. A comparison of six different ballast water treatment systems based on UV radiation, electrochlorination and chlorine dioxide. Environ. Technol. 36, 2094–104. doi:10.1080/09593330.2015.1021858

Veldhuis, M.J.W., Fuhr, F., Boon, J.P., Hallers-tjabbers, C.C. Ten, 2006. Treatment of Ballast Water ; How to Test a System with a Modular Concept ? 27, 909–921. doi:10.1080/09593332708618701

Veldhuis, M.J.W., Kraay, G.W., 2000. Application of flow cytometry in marine phytoplankton research : current applications and future perspectives * 64, 121–134.

Wollschläger, J., Nicolaus, A., Wiltshire, K.H., Metfies, K., 2014. Assessment of North Sea phytoplankton via molecular sensing: A method evaluation. J. Plankton Res. 36, 695–708. doi:10.1093/plankt/fbu003